\documentclass[11pt]{article}
\usepackage{graphicx}

\newcommand{\BABARPubYear}    {06}

\newcommand{\BABARProcNumber} {089}
\newcommand{\SLACPubNumber} {12145}

\RequirePackage{xspace}
\def\lbabar{\mbox{{\large\sl B}\hspace{-0.4em} {\normalsize\sl A}\hspace{-0.03em}{\large\sl B}\hspace{-0.4em} {\normalsize\sl A\hspace{-0.02em}R}}}
\usepackage{relsize}
\def\babar{\mbox{\slshape B\kern-0.1em{\smaller A}\kern-0.1em
    B\kern-0.1em{\smaller A\kern-0.2em R}}}
\def\fbar  {\ensuremath{\overline f}\xspace}

\def\ccbar {\ensuremath{c\overline c}\xspace}

\def\t     {\ensuremath{t}\xspace}

\newcommand{\btoccbars}{\ensuremath{b\rightarrow c {\overline c} s}\xspace}

\newcommand{\Acp}{\ensuremath{A_{\CP}}}\xspace
\newcommand{\Bcp}{\ensuremath{B_{\CP}}}\xspace
\newcommand{\Btag}{\ensuremath{B_{tag}}}\xspace
\newcommand{\tcp}{\ensuremath{t_{\CP}}}\xspace
\newcommand{\ttag}{\ensuremath{t_{tag}}}\xspace

\def\piz   {\ensuremath{\pi^0}\xspace}

\def\pip   {\ensuremath{\pi^+}\xspace}
\def\pim   {\ensuremath{\pi^-}\xspace}

\def\Kbar  {\kern 0.2em\overline{\kern -0.2em K}{}\xspace}

\def\Kz    {\ensuremath{K^0}\xspace}
\def\Kzb   {\ensuremath{\Kbar^0}\xspace}
\def\KzKzb {\ensuremath{\Kz \kern -0.16em \Kzb}\xspace}
\def\Kp    {\ensuremath{K^+}\xspace}
\def\Km    {\ensuremath{K^-}\xspace}

\def\KpKm  {\ensuremath{\Kp \kern -0.16em \Km}\xspace}
\def\KS    {\ensuremath{K^0_{\scriptscriptstyle S}}\xspace} 
\def\KL    {\ensuremath{K^0_{\scriptscriptstyle L}}\xspace} 
\def\Kstarz  {\ensuremath{K^{*0}}\xspace}

\def\Kstar   {\ensuremath{K^*}\xspace}

\newcommand{\hz}{\ensuremath{h^0}\xspace}

\def\Dbar    {\kern 0.2em\overline{\kern -0.2em D}{}\xspace}

\def\Dz      {\ensuremath{D^0}\xspace}
\def\Dzb     {\ensuremath{\Dbar^0}\xspace}
\def\DzDzb   {\ensuremath{\Dz {\kern -0.16em \Dzb}}\xspace}
\def\Dp      {\ensuremath{D^+}\xspace}
\def\Dm      {\ensuremath{D^-}\xspace}

\def\DpDm    {\ensuremath{\Dp {\kern -0.16em \Dm}}\xspace}

\def\Dstarz  {\ensuremath{D^{*0}}\xspace}

\def\Dstarp  {\ensuremath{D^{*+}}\xspace}
\def\Dstarm  {\ensuremath{D^{*-}}\xspace}

\def\B       {\ensuremath{B}\xspace}
\def\Bbar    {\kern 0.18em\overline{\kern -0.18em B}{}\xspace}

\def\BB      {\ensuremath{B\Bbar}\xspace} 
\def\Bz      {\ensuremath{B^0}\xspace}
\def\Bzb     {\ensuremath{\Bbar^0}\xspace}
\def\BzBzb   {\ensuremath{\Bz {\kern -0.16em \Bzb}}\xspace}
\def\Bu      {\ensuremath{B^+}\xspace}
\def\Bub     {\ensuremath{B^-}\xspace}

\def\BpBm    {\ensuremath{\Bu {\kern -0.16em \Bub}}\xspace}

\def\BorBbar    {\kern 0.18em\optbar{\kern -0.18em B}{}\xspace}
\def\DorDbar    {\kern 0.18em\optbar{\kern -0.18em D}{}\xspace}
\def\KorKbar    {\kern 0.18em\optbar{\kern -0.18em K}{}\xspace}

\def\jpsi     {\ensuremath{{J\mskip -3mu/\mskip -2mu\psi\mskip 2mu}}\xspace}

\def\psitwos  {\ensuremath{\psi{(2S)}}\xspace}

\def\etac     {\ensuremath{\eta_c}\xspace}

\def\chicone  {\ensuremath{\chi_{c1}}\xspace}

\mathchardef\Upsilon="7107
\def\Y#1S{\ensuremath{\Upsilon{(#1S)}}\xspace}% no space before {...}!

\def\FourS {\Y4S}

\mathchardef\Deltares="7101
\mathchardef\Xi="7104
\mathchardef\Lambda="7103
\mathchardef\Sigma="7106
\mathchardef\Omega="710A

\def\Deltabar{\kern 0.25em\overline{\kern -0.25em \Deltares}{}\xspace}
\def\Lbar{\kern 0.2em\overline{\kern -0.2em\Lambda\kern 0.05em}\kern-0.05em{}\xspace}
\def\Sigbar{\kern 0.2em\overline{\kern -0.2em \Sigma}{}\xspace}
\def\Xibar{\kern 0.2em\overline{\kern -0.2em \Xi}{}\xspace}
\def\Obar{\kern 0.2em\overline{\kern -0.2em \Omega}{}\xspace}
\def\Nbar{\kern 0.2em\overline{\kern -0.2em N}{}\xspace}
\def\Xb{\kern 0.2em\overline{\kern -0.2em X}{}\xspace}

\def\bpsikst    {\ensuremath{\Bz \to \jpsi \Kstar}\xspace}

\def\Bztodstdstks {\ensuremath{\Bz \to \Dstarp \Dstarm \KS}\xspace}
\def\Bztodsthz  {\ensuremath{\Bz \to \Dstarz \hz}\xspace}
\def\DsChunhui  {\ensuremath{D^-_{s1}(2536)}\xspace}
\def\Dssone     {\ensuremath{D^+_{s1}}\xspace}
\def\dstarelnubulo{\ensuremath{\Dstarm\ell^+\nu_{\ell}}\xspace}
\newcommand{\ImLambda}{\ensuremath{\rm Im\, \vert\lambda\vert}}
\newcommand{\z}{\ensuremath{{\mathsf z}}\xspace}
\newcommand{\Imz}{\ensuremath{\rm Im\, \z}}
\newcommand{\ImzZero}{\ensuremath{\rm Im\, \z_0}}
\newcommand{\ImzOne}{\ensuremath{\rm Im\, \z_1}}
\newcommand{\Rez}{\ensuremath{\rm Re\, \z}}
\newcommand{\dG}{\ensuremath{ \Delta \Gamma }}

\newcommand{\dGRezZero}{\ensuremath{\dG\cdot\Rez_{0}}}
\newcommand{\dGRezOne}{\ensuremath{\dG\cdot\Rez_{1}}}
\def\Jzero      {\ensuremath{J_{0}}}
\def\Jsone      {\ensuremath{J_{s1}}}
\def\Jstwo      {\ensuremath{J_{s2}}}
\def\Jc         {\ensuremath{J_{c}\xspace}}

\def\Asame      {\ensuremath{A_{\T/\CP}}}\xspace
\def\Aopp       {\ensuremath{A_{\T/\CPT}}}\xspace

\newcommand{\tev}{\ensuremath{\mathrm{\,Te\kern -0.1em V}}\xspace}
\newcommand{\gev}{\ensuremath{\mathrm{\,Ge\kern -0.1em V}}\xspace}
\newcommand{\mev}{\ensuremath{\mathrm{\,Me\kern -0.1em V}}\xspace}
\newcommand{\kev}{\ensuremath{\mathrm{\,ke\kern -0.1em V}}\xspace}
\newcommand{\ev}{\ensuremath{\mathrm{\,e\kern -0.1em V}}\xspace}
\newcommand{\gevc}{\ensuremath{{\mathrm{\,Ge\kern -0.1em V\!/}c}}\xspace}
\newcommand{\mevc}{\ensuremath{{\mathrm{\,Me\kern -0.1em V\!/}c}}\xspace}
\newcommand{\gevcc}{\ensuremath{{\mathrm{\,Ge\kern -0.1em V\!/}c^2}}\xspace}
\newcommand{\mevcc}{\ensuremath{{\mathrm{\,Me\kern -0.1em V\!/}c^2}}\xspace}
\def\mus  {\ensuremath{\rm \,\mus}\xspace}

\def\ps   {\ensuremath{\rm \,ps}\xspace}

\def\mus        {\ensuremath{\,\mu{\rm s}}\xspace}    %% microsecond
\def\ps         {\ensuremath{{\rm \,ps}}\xspace}  %% picosecond

\def\to                 {\ensuremath{\rightarrow}\xspace}

\def\pep2{PEP-II}

\def\gsim{{~\raise.15em\hbox{$>$}\kern-.85em
          \lower.35em\hbox{$\sim$}~}\xspace}
\def\lsim{{~\raise.15em\hbox{$<$}\kern-.85em
          \lower.35em\hbox{$\sim$}~}\xspace}

\def\CP                {\ensuremath{C\!P}\xspace}
\def\CPT               {\ensuremath{C\!PT}\xspace} % Looks better without \!
\def\C       {\ensuremath{C}\xspace}
\def\S       {\ensuremath{S}\xspace}

\def\T       {\ensuremath{T}\xspace}

\def\rhobar {\ensuremath{\overline \rho}\xspace}
\def\etabar {\ensuremath{\overline \eta}\xspace}
\def\stwob{\ensuremath{\sin\! 2 \beta   }\xspace}
\def\ctwob{\ensuremath{\cos\! 2 \beta   }\xspace}

\def\deltat{\ensuremath{{\rm \Delta}t}\xspace}
\def\deltamd{\ensuremath{{\rm \Delta}m_d}\xspace}

\xspace

\newcommand{\jprlBase}       {Phys.\ Rev.\ Lett.\xspace}
\newcommand{\jprBase}        {Phys.\ Rev.\xspace}
\newcommand{\jplBase}        {Phys.\ Lett.\xspace}
\newcommand{\nimBaseA}       {Nucl.\ Instr.\ Methods Phys.\ Res., Sect.\ A\xspace}

\newcommand{\npBase}         {Nucl.\ Phys.\xspace}

\newcommand{\nima}      [1]  {\nimBaseA~{\bf #1}}

\newcommand{\npb}       [1]  {\npBase\ B~{\bf #1}}

\newcommand{\plb}       [1]  {\jplBase\ B~{\bf #1}}

\newcommand{\jprl}      [1]  {\jprlBase\ {\bf #1}}
\newcommand{\jprd}      [1]  {\jprBase\ D~{\bf #1}}
\def\jetset74   {\mbox{\tt Jetset \hspace{-0.5em}7.\hspace{-0.2em}4}\xspace}

\def\effectiveeta{0.504 \pm 0.033}

\long\def\inst#1{\par\nobreak\kern 4pt\nobreak
    {\it #1}\par\vskip 10pt plus 3pt minus 3pt}

\RequirePackage{xspace}

\begin{document}
{\pagestyle{empty}

\begin{flushright}
SLAC-PUB-\SLACPubNumber \\
\babar-PROC-\BABARPubYear/\BABARProcNumber \\
%hep-ex/\LANLNumber \\
\end{flushright}

\par\vskip 4cm

% Title of the paper
\begin{center}
\Large \bf {\boldmath{\CP}} Violation measurements in B\to\ charm decays at \babar.
\end{center}
\bigskip

\begin{center}
\large 
Katherine George\\
Queen Mary, University of London.\\
Department of Physics, Mile End Road, London E1 4NS, United~Kingdom.\\
(representing the \lbabar\ Collaboration)
\end{center}
\bigskip \bigskip

%----------------------
% Abstract
%----------------------
\begin{center}
\large \bf Abstract
\end{center}
This article summarises measurements of time-dependent $\CP$ asymmetries in decays of neutral \B\ mesons to charm final
states using data collected by the \babar\ detector at the PEP-II asymmetric-energy \B\-factory.
All results are preliminary unless otherwise stated.}

\vfill
\begin{center}
Contributed to the Proceedings of ICHEP'06 - XXXIII International Conference on \\
High Energy Physics, Moscow, Russia. July 26 - August 2, 2006. 
\end{center}

\vspace{1.0cm}
\begin{center}
{\em Stanford Linear Accelerator Center, Stanford University, 
Stanford, CA 94309} \\ \vspace{0.1cm}\hrule\vspace{0.1cm}
Work supported in part by Department of Energy contract DE-AC02-76SF00515.
\end{center}

%-----------
% Sections
%-----------
%----------------------
\section{Introduction}
%----------------------
%Charge conjugation-parity (\CP) violation in the $B$ meson system has been
%established by the \babar~\cite{ref:babar-stwob-prl} and Belle~\cite{ref:belle-stwob-prl} collaborations.
The Standard Model (SM) of particle physics describes \CP\ violation
as a consequence of a complex phase in the three-generation Cabibbo-Kobayashi-Maskawa (CKM) quark-mixing
matrix~\cite{ref:ckm}. In this framework, measurements of \CP\ asymmetries in
the proper-time distribution of neutral $B$ decays to \CP\ eigenstates containing a charmonium and $K^{0}$ meson provide
a direct measurement of $\stwob$~\cite{ref:BCP}. The unitarity triangle angle $\beta$ is
$\arg \left[\, -V_{\rm cd}^{}V_{\rm cb}^* / V_{\rm td}^{}V_{\rm tb}^*\, \right]$ where
the $V_{ij}$ are CKM matrix elements.

The BaBar detector~\cite{ref:babar} is located at the SLAC PEP-II $e^+e^-$ asymmetric energy \B -factory.
Its program includes the measurement of the angle $\beta$ through the measurement of time-dependent
$\CP$-asymmetries, $\Acp$. At the $\Upsilon(4S)$ resonance, $\Acp$ is extracted from the distribution of the
difference of the proper decay times, $\t \equiv \tcp - \ttag$, where
$\tcp$ refers to the decay time of the signal \B\ meson ($\Bcp$) and $\ttag$ refers to the
decay time of the other \B\ meson in the event ($\Btag$). The decay products of $\Btag$ are
used to identify its flavor at its decay time.
%----------------------------
\begin{eqnarray}
\Acp(t) & \equiv & \frac{N(\Bzb(t)\to f) - N(\Bz(t)\to f)} {N(\Bzb(t)\to f) + N(\Bz(t)\to f)} \nonumber \\
        &&{} = \S \sin(\deltamd{t}) - \C \cos(\deltamd{t}),\nonumber
\label{eq:timedependence}
\end{eqnarray}
%----------------------------
where $N(\Bzb(t)\to f)$ is the number of \Bzb\ that decay into the $CP$-eigenstate $f$ after a time $t$ 
and $\deltamd$ is the difference between the \B\ mass eigenstates. 
The sinusoidal term describes interference between mixing and decay and the cosine term is the
direct \CP\ asymmetry. \S and \C are functions of the parameter $\lambda$ and are defined as follows:
%----------------------------
% lambda, S and C equations
%----------------------------
\begin{eqnarray}
\S = {\frac{{2\cdot\ImLambda}}{{1+{\vert\lambda\vert}^2}}}, \nonumber \\
\C = {\frac{{1-{\vert\lambda\vert}^2}}{{1+{\vert\lambda\vert}^2}}}, \nonumber \\
\lambda  =  {\frac{q}{p}}\cdot\frac{A(\Bzb(t)\to \fbar)}{A(\Bz(t)\to f)}.
\label{eq:sandc}
\end{eqnarray}
%----------------------------
In Eq.~\ref{eq:sandc}, $A(\Bzb(t)\to \fbar)$ ($A(\Bz(t)\to f)$) is the decay amplitude of 
$\Bzb$ ($\Bz$) to the final state $\fbar$ ($f$). 
The physical states (solutions of the complex effective Hamiltonian for the \Bz-\Bzb system) can be written in
terms of the parameters $p$, $q$ and $\z$~\cite{ref:schneider}:
%----------------------------
\begin{eqnarray}
\vert{\B_L}\rangle & = & p\sqrt{1 - \z}\vert{\Bz}\rangle + q\sqrt{1 + \z}\vert{\Bzb}\rangle,\nonumber \\\newline
\vert{\B_H}\rangle & = & p\sqrt{1 + \z}\vert{\Bz}\rangle - q\sqrt{1 - \z}\vert{\Bzb}\rangle,\nonumber
\end{eqnarray}
%----------------------------
where $H$ and $L$ denote the Heavy and Light mass eigenstates. 
Under \CPT symmetry, the complex parameter \z vanishes. 
\T invariance implies $\vert{q/{p}}\vert$ = 1 and \CP invariance requires both $\vert{q/{p}}\vert$ = 1 and \z = 0.
In this article the current status of measurements of \CP\ violation in $\B\rightarrow$ charm decays and 
studies of searches for \T, \CPT and \CP violation in $\Bz$-$\Bzb$ mixing are presented. 
All results are preliminary unless otherwise stated.
%--------------------------------------------------------------------------------------------------------------
\section{$\btoccbars$ decay modes}
\label{sec:jpsiks}
\begin{figure}
\begin{center}
%\vspace{1.0cm}
\begin{center}
\scalebox{0.7}{\includegraphics{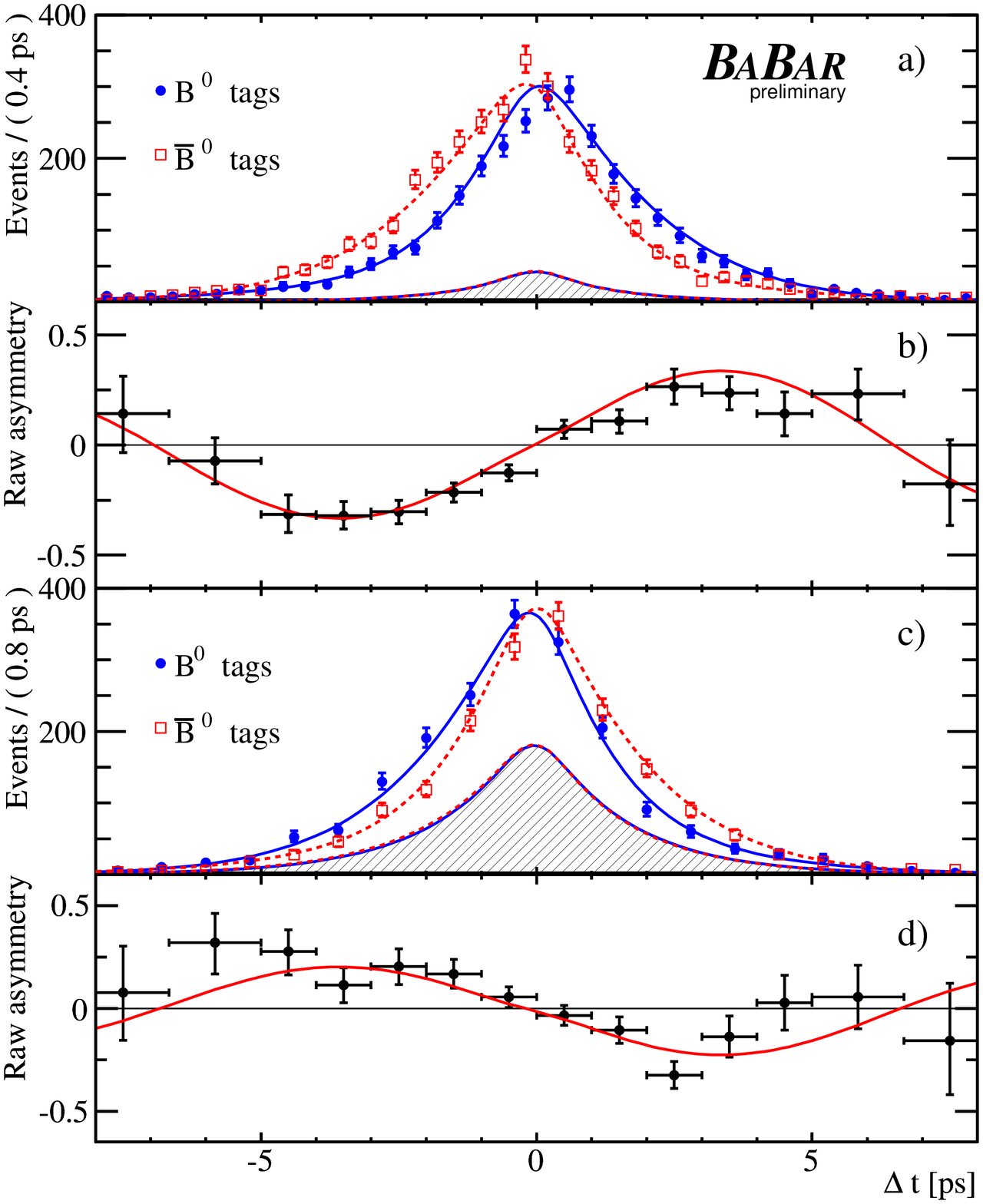}}
\end{center}
%\vspace{1.0cm}
\caption{
a) Number of $\eta_f=-1$ candidates ($J/\psi \KS$, $\psi(2S) \KS$, $\chicone \KS$, and $\eta_c \KS$)
in the signal region with a \Bz tag $N_{\Bz }$ and with a \Bzb tag $N_{\Bzb}$, and 
b) the raw asymmetry $(N_{\Bz}-N_{\Bzb})/(N_{\Bz}+N_{\Bzb})$, as functions of \deltat.
Figs. c) and d) are the corresponding plots for the $\eta_f=+1$ mode $J/\psi \KL$.  
%All plots exclude othertag-tagged events. 
The solid (dashed) curves represent the fit projections in \deltat for \Bz (\Bzb) tags. 
The shaded regions represent the estimated background contributions.
\label{fig:sin2BAsymm}}
\end{center}
\end{figure}

%--------------------------------------------------------------------------------------------------------------
The determination of $\beta$ from $\btoccbars$ decay modes currently provides the most stringent constraint
on the unitarity triangle. 
For these decay modes, the \CP\ violation parameters in Eq.~\ref{eq:sandc} are 
$\S_{\btoccbars} = -\eta_{f}\stwob$ and $\S_{\btoccbars}$ = 0, where 
$\eta_{f}$ is $-$1 for ($\ccbar$)$\KS$ decays (e.g. $\jpsi\KS$, $\psitwos\KS$, $\chicone\KS$, $\eta_c \KS$~\cite{ref:charge}) 
and $\eta_{f}$ is $+$1 for the ($\ccbar$)$\KL$ (e.g. $\jpsi\KL$) state.
We use the value $\eta_f = \effectiveeta$ for the 
$\jpsi\Kstarz (\Kstarz \to \KS\piz)$ final state since it can be both \CP even and \CP odd  
due to the presence of even and odd orbital angular momentum contributions~\cite{ref:rperp}.
These modes have been used to measure $\stwob$ using 348 M $\BB$ pairs~\cite{ref:sin2b},
where an improved event reconstruction has been applied to the complete dataset, and a
new $\eta_c \KS$ event selection has been developed based on the Dalitz structure of the 
$\etac \to \KS \Kp \pim$ decay. We measure~\cite{ref:errors}
%----------------------------
\begin{eqnarray}
\stwob & = & 0.715 \pm 0.034 \pm 0.019, \nonumber \\  
\vert\lambda\vert & = & 0.932 \pm 0.026 \pm 0.017 \nonumber
\end{eqnarray}
%----------------------------
which is in agreement with SM expectations. 
Figure~\ref{fig:sin2BAsymm} shows the \deltat distributions and asymmetries in yields between \Bz tags and \Bzb tags for the
$\eta_f=-1$ and $\eta_f = +1$ samples as a function of \deltat, overlaid with the projection of the likelihood fit result.
%--------------------------------------
% Figure : Deltat distributions etc ...
%--------------------------------------
%--------------------------------------
%-------------------------------------------------------------------------------------------
\section{$\ctwob$ measurements}
\label{sec:cos2beta}
%-------------------------------------------------------------------------------------------
The analysis of $\btoccbars$ decay modes imposes a constraint on $\stwob$ only, leading to a four-fold
ambiguity in the determination of $\beta$. This ambiguity can leave possible new physics undetected
even with very high precision measurements of $\stwob$. Additional constraints are obtained from the 
ambiguity-free measurement of $\ctwob$ using the angular and time-dependent asymmetry in $\bpsikst$ decays
and the time-dependent Dalitz plot analyses of $\Bztodsthz$ and $\Bztodstdstks$.
The $\bpsikst$ analysis is published in Ref.~\cite{ref:cos2beta}.  

A model-independent measurement of $\ctwob$ in $\Bztodsthz$ decays has been made using a time-dependent Dalitz plot
analysis of $\Dz\rightarrow\KS\pip\pim$, where $\hz$ is a light neutral meson such as $\piz$, $\eta$, $\eta^\prime$ or 
$\omega$~\cite{ref:cheng}.
The strong phase variation on the $\Dz\rightarrow\KS\pip\pim$ Dalitz plot allows access to the angle $\beta$ 
with only a two-fold ambiguity (${\beta + \pi}$)~\cite{ref:bondar}. Using 311 M \BB pairs, the following values 
of the \CP\ asymmetry parameters are extracted: 
%-----------------------------------
\begin{eqnarray}
\ctwob & = & 0.54 \pm 0.54 \pm 0.08 \pm 0.18, \nonumber \\
\stwob & = & 0.45 \pm 0.36 \pm 0.05 \pm 0.07, \nonumber \\
\vert\lambda\vert & = & 0.975 ^{+0.093}_{-0.085} \pm 0.012 \pm 0.002,\nonumber
\end{eqnarray}
%-----------------------------------
where in addition to the statistical and systematic errors, there are also uncertainties from the signal Dalitz model. 
Assuming that $\stwob$ takes the same value as the $\btoccbars$ decay average in Ref.~\cite{ref:hfag} 
%and that there is no \CP\ violation in \B\ decays, 
and that there is no \CP\ violation in $\Bz$-$\Bzb$ mixing,
a parameterised Monte Carlo method based on the observed data finds that 
these measurements favour the solution of $\beta$ = 22$^\circ$ over 68$^\circ$ at an 87$\%$ confidence level.  

A study of the decay $\Bztodstdstks$ has been made using 230 M \BB pairs~\cite{ref:poireau}.
The branching fraction ${\cal{B}}(\Bztodstdstks) = (4.4 \pm 0.4 \pm 0.7) \times 10^{-3}$
has been measured and evidence found for the decay $\Bz\rightarrow\Dstarp\DsChunhui\KS$ with a 4.6$\sigma$ statistical significance. 
The time-dependent decay rate asymmetry of $\Bztodstdstks$ can be written in terms of the parameters 
$\Jzero$, $\Jsone$, $\Jstwo$ and $\Jc$ which are integrals over the half Dalitz space of the decay amplitudes
of $\Bz\rightarrow\Dstarp\Dstarm\KS$ and $\Bzb\rightarrow\Dstarp\Dstarm\KS$~\cite{ref:browder}. 
The fits to the data yield:
%-----------------------------------
\begin{eqnarray}
{\frac{\Jc}{\Jzero}}           & = & 0.76 \pm 0.18 \pm 0.07, \nonumber \\
{\frac{2\Jsone}{\Jzero}}\stwob & = & 0.10 \pm 0.24 \pm 0.06, \nonumber \\
{\frac{2\Jstwo}{\Jzero}}\ctwob & = & 0.38 \pm 0.24 \pm 0.05. \nonumber
\end{eqnarray}
%-----------------------------------
The measured value of ${\Jc}/{\Jzero}$ is significantly different from zero, which, according 
to Ref.~\cite{ref:browder}, may indicate that there is a sizeable broad resonant contribution to the 
decay $\Bztodstdstks$ from an unknown $\Dssone$ state with an unknown width. Under this assumption
then the measured value of ${2\Jstwo}/{\Jzero}$ implies that the sign of $\ctwob$ is preferred to be 
positive at a 94$\%$ confidence level. 

\begin{figure}[!htb]
\begin{center}
\begin{center}
\scalebox{0.5}{\includegraphics{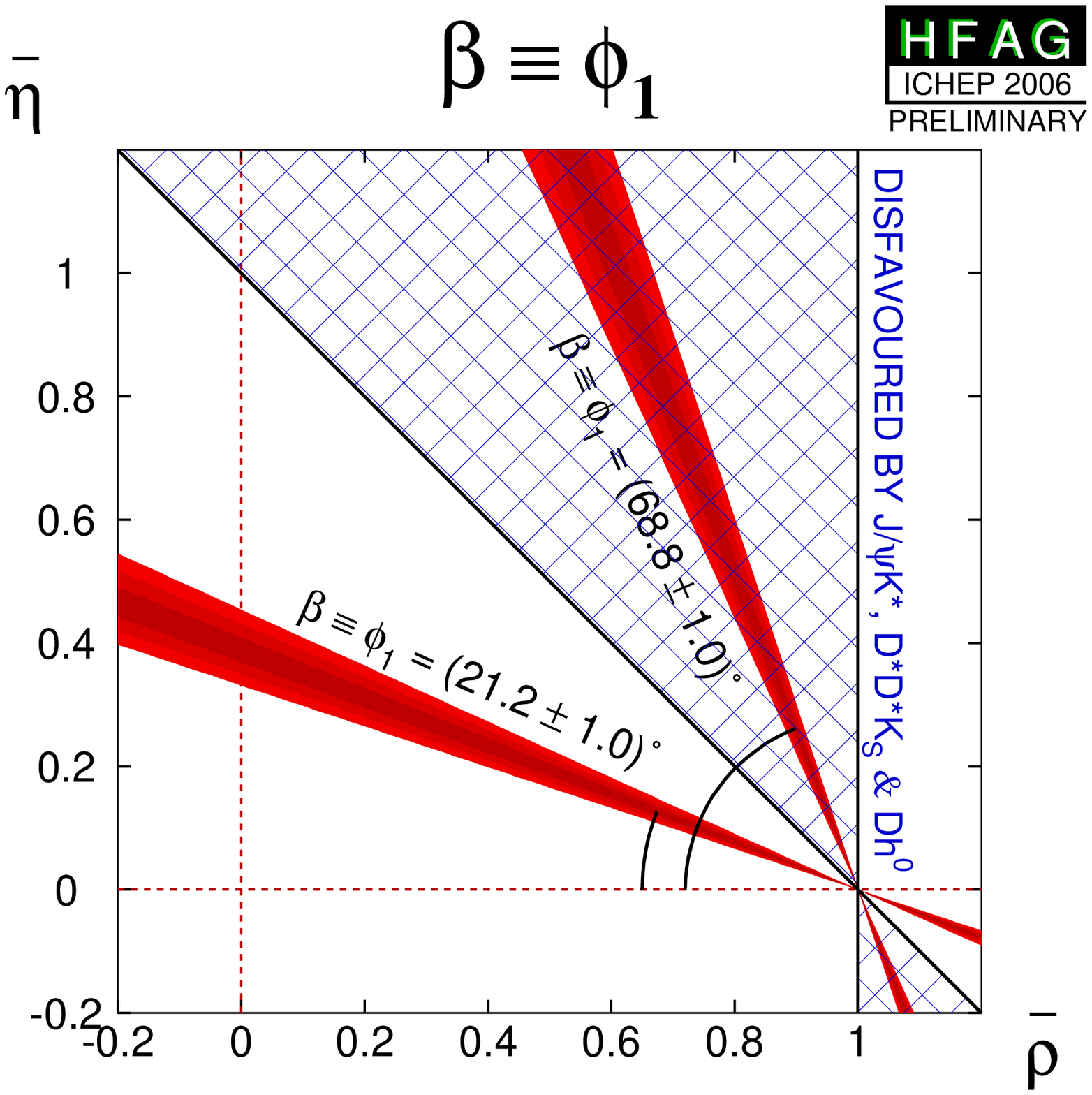}}
\end{center}
\caption{Constraint on $\beta$ in the $\rhobar$-$\etabar$ plane, obtained from the analysis of $\btoccbars$ decays, 
the angular analysis of $\bpsikst$ and the time-dependent Dalitz plot analyses of $\Bztodsthz$ and $\Bztodstdstks$. 
The hatched area corresponding to the solution $\beta = $68.8$\pm$1.0$^\circ$ where $\ctwob$ is negative, is strongly disfavoured.
\label{fig:hfag}}
\end{center}
\end{figure}

Figure~\ref{fig:hfag} illustrates the combined constraint on $\beta$ in the $\rhobar$-$\etabar$ plane from
the Belle and \babar\ $\btoccbars$, $\bpsikst$, $\Bztodsthz$ and $\Bztodstdstks$ analyses~\cite{ref:hfag}.
%---------------------------------------------------------------------
% HFAG : etabar - rhobar plane
%---------------------------------------------------------------------
%--------------------------------------
%--------------------------------------------------------------------
%\section{$\Bz\Bzb$ mixing measurements}
\section{Studies of \T, \CPT and \CP violation in $\Bz$-$\Bzb$ mixing.}
\label{sec:mixing}
\begin{figure}[!htb]
\begin{center}
\scalebox{0.7}{\includegraphics[height=9cm]{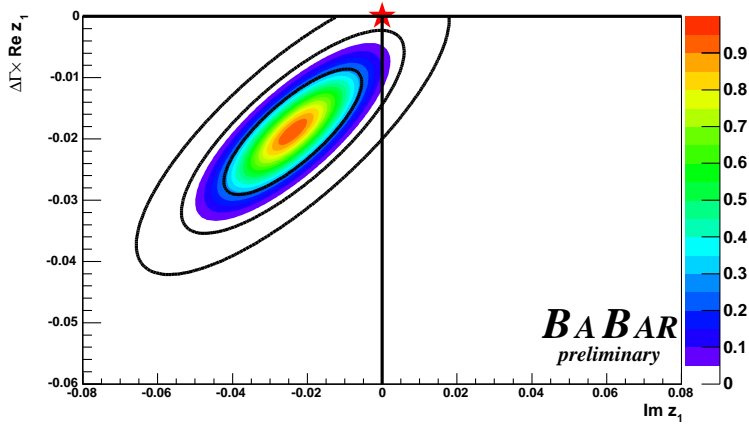}}
\caption{Confidence level contours for the parameters $\ImzOne$ and $\dGRezOne$. 
The line contours indicate $1\sigma$, $2\sigma$, and $3\sigma$ significance. 
The star at $\ImzOne = \dGRezOne = 0$ indicates the condition for no sidereal-time dependence in \z.}
\label{fig:stoker}
\end{center}
\end{figure}

%--------------------------------------------------------------------
Inclusive dilepton events, where both \B mesons decay semileptonically represent 4$\%$ of all
$\FourS\to\BzBzb$ decays and provide a very large sample with which to study \T, \CPT and \CP
violation in $\Bz$-$\Bzb$ mixing. 
The same-sign dilepton asymmetry $\Asame$ between the two oscillation probabilities 
$P(\Bzb\to\Bz)$ and $P(\Bz\to\Bzb)$ is sensitive to $\vert{q/p}\vert$ and probes both \T and \CP symmetries. 
The opposite-sign dilepton asymmetry $\Aopp$ compares the probabilities $P(\Bz\to\Bz)$ and $P(\Bzb\to\Bzb)$
and probes \CPT and \CP violation. It is sensitive to the product $\Delta\Gamma\cdot\Rez$
where $\Delta\Gamma$ is the difference between the decay rates of the neutral \B mass eigenstates.
The result published in Ref.~\cite{ref:legendre} uses a sample of 232 M \BB pairs to measure 
the \T and \CP violation parameter
%-----------------------------------
\begin{eqnarray}
{\vert{q/p}\vert} - 1  =  (-0.8 \pm 2.7 \pm 1.9)\times{10}^{-3}\nonumber
\end{eqnarray}
%-----------------------------------
and the \CPT and \CP parameters
%-----------------------------------
\begin{eqnarray}
\Imz & = & (-13.9 \pm 7.3 \pm 3.2)\times{10}^{-3}, \nonumber \\
\Delta\Gamma \cdot \Rez & = & (-7.1 \pm 3.9 \pm 2.0)\times{10}^{-3}\:ps^{-1}.\nonumber
\end{eqnarray}
%-----------------------------------
The statistical correlation between the measurements of $\Imz$ and $\Delta\Gamma\cdot\Rez$ is 76$\%$. 
A search is then made for time-dependent variations in the complex \CPT parameter 
$\z = \z_0 + \z_1\cos{(\Omega\hat{t} + \phi)}$ where $\Omega$ is the
Earth's sidereal frequency and $\hat{t}$ is sidereal time~\cite{ref:stoker}.  We measure:
%-----------------------------------
\begin{eqnarray}
\ImzZero & = & (-14.1 \pm 7.3 \pm 2.4)\times 10^{-3},\nonumber \\
\dGRezZero & = & (-7.2 \pm 4.1 \pm 2.1)\times 10^{-3}\ps^{-1},\nonumber \\
\ImzOne & = & (-24.0 \pm 10.7 \pm 5.9)\times 10^{-3}, \nonumber \\
\dGRezOne & = & (-18.8 \pm 5.5 \pm 4.0)\times 10^{-3}\ps^{-1}.\nonumber 
\end{eqnarray}
%-----------------------------------
The statistical correlation between the measurements of \ImzZero\ and \dGRezZero\ is 76\%; 
and between \ImzOne\ and \dGRezOne\ is 79\%. 
Figure~\ref{fig:stoker} shows confidence level contours for the
parameters \ImzOne\ and \dGRezOne\ including both statistical and systematic
errors. A significance of $2.2\sigma$ is found for periodic variations in the \CPT
violation parameter \z\ at the sidereal frequency, characteristic of
Lorentz violation.
%----------------------------------------------------------
% Figure : Lorentz violation confidence level contour plot 
%----------------------------------------------------------
%----------------------------------------------------------

A measurement of the parameter $\vert{q/p}\vert$ has also been made using the
partial reconstruction of one of the $B$ mesons in the semileptonic channel 
\dstarelnubulo, where only the hard lepton and the soft pion from the 
$\Dstarm \rightarrow \Dzb \pi^-$ decay are reconstructed~\cite{ref:gaz}. 
A data sample of 220 M \BB pairs are used. We measure
%-----------------------------------
\begin{eqnarray}
\vert{q/p} \vert -1 = (6.5 \pm 3.4 \pm 2.0)\times 10^{-3}\nonumber 
\end{eqnarray}
%-----------------------------------
which is consistent with SM expectations.

%---------------------------------------------------
\section{Conclusion}
\label{sec:conclusion}
%---------------------------------------------------
An improved measurement of $\stwob$ has been made using $\Bz \rightarrow$ charmonium + $\Kz$ decays.
This is consistent with SM expectations and continues to provide the most stringent constraint
on the unitarity triangle.
Analysis of the $\btoccbars$, $\bpsikst$, $\Bztodsthz$ and $\Bztodstdstks$ modes indicate that the solution 
$\beta =$ 21.1 $\pm$ 1.0$^\circ$ is strongly preferred.
The measurements of $\vert{q/p} \vert$ from analyses of inclusive dilepton and $\dstarelnubulo$ events are in agreement with SM predictions.

%=============================
% BIBLIOGRAPHY
%=============================


\begin{thebibliography}{99}
% [1] CKM matrix
\bibitem{ref:ckm} 
N.~Cabibbo, \jprl{10}, 531 (1963);\newline  
M.~Kobayashi and T.~Maskawa, Prog.\ Th.\ Phys.\ {\bf 49}, 652 (1973).

% [2] Direct measurement of sin(2\beta)
\bibitem{ref:BCP}
A.B.~Carter and A.I.~Sanda, \jprd{23}, 1567 (1981);\newline 
I.I.~Bigi and A.I.~Sanda, \npb{193}, 85 (1981).

% [3] NIM detector performance paper
\bibitem{ref:babar}
\babar\ Collaboration, B.\ Aubert {\em et al.},
\nima{479}, 1 (2002).

% [4] B0 - anti-B0 mixing section for the 2006 PDG
\bibitem{ref:schneider}
O.~Schneider, arXiv:hep-ex/0606040 (2006).

% [5] Charge-conjugate reactions
\bibitem{ref:charge}
Charge-conjugate reactions are included implicitly unless otherwise specified.

% [6] Measurement of Decay Amplitudes of \B \to (\ccbar) \Kstar with an angular analysis, for (\ccbar)=\jpsi, \psitwos and \chicone
\bibitem{ref:rperp}
\babar\ Collaboration, B.\ Aubert {\em et al.}, arXiv:hep-ex/0607081 (2006).

% [7] Improved measurements of time-dependent CP violation in B to ccbar K decays
\bibitem{ref:sin2b}   
\babar\ Collaboration, B.\ Aubert {\em et al.}, arXiv:hep-ex/0607107 (2006).

% [8] Systematic and statistical errors
\bibitem{ref:errors}
Unless otherwise stated, all results are quoted with the first error being statistical and the second systematic.

% [9] Run 1 + 2 cos2B from J/psi Kstar angular analysis
\bibitem{ref:cos2beta}
\babar\ Collaboration, B.\ Aubert {\em et al.}, \jprd{71} 032005 (2005).

% [10] Measurements of the CKM angle beta with B0 to D(*)0h0 decays using a Dalitz analysis of D0 to Ks pi+ pi- 
\bibitem{ref:cheng} 
\babar\ Collaboration, B.\ Aubert {\em et al.}, arXiv:hep-ex/0607105 (2006).

% [11] A method to measure phi(1) using anti-B0 --> D h0 with multibody D decay 
\bibitem{ref:bondar}
A.~Bondar {\em et al.}, \plb{624} (2005).

% [12] HFAG
\bibitem{ref:hfag}
Heavy Flavor Averaging Group: http://www.slac.stanford.edu/xorg/hfag 

% [13] Branching fraction measurement and time-dependent analysis for the decay of B0 to D*+D*-Ks
\bibitem{ref:poireau}
\babar\ Collaboration, B.\ Aubert {\em et al.}, arXiv:hep-ex/0608016 (2006).

% [14] Theoretical assumptions for D*+D*-Ks analysis
\bibitem{ref:browder}
T.E.~Browder {\em et al.}, \jprd{61}, 054009 (2000).

% [15] Search for T, CP and CPT Violation in B0-B0bar mixing with inclusive dileption events
\bibitem{ref:legendre} 
\babar\ Collaboration, B.\ Aubert {\em et al.}, \jprl{96}, 251802 (2006).

% [16] Search for T, CP and CPT Violation in B0-B0bar mixing with inclusive dileption events
\bibitem{ref:stoker} 
\babar\ Collaboration, B.\ Aubert {\em et al.}, arXiv:hep-ex/0607103 (2006).

% [17] Study of CP Violation with partially reconstructed B0 to D*lnu decays
\bibitem{ref:gaz}
\babar\ Collaboration, B.\ Aubert {\em et al.}, arXiv:hep-ex/0607091 (2006).

\end{thebibliography}
\end{document}